\documentclass[aps,twocolumn,superscriptaddress,prl]{revtex4}
\usepackage{amssymb}
\usepackage{amsmath}
\usepackage{graphicx}
\usepackage[usenames]{color}

\begin{document}

\title{Conditional operation of a spin qubit}
\author{I.\ van\ Weperen}
\affiliation{Department of Physics, Harvard University, Cambridge, Massachusetts 02138, USA}
\affiliation{Kavli Institute of Nanoscience, Delft University of Technology, 2600 GA Delft, The Netherlands}
\author{B.\ D.\ Armstrong}
\affiliation{Department of Physics, Harvard University, Cambridge, Massachusetts 02138, USA}
\author{E.\ A.\ Laird}
\affiliation{Department of Physics, Harvard University, Cambridge, Massachusetts 02138, USA}
\altaffiliation{present address: Kavli Institute of Nanoscience, Delft University of Technology, 2600 GA Delft, The Netherlands}
\author{J.\ Medford}
\affiliation{Department of Physics, Harvard University, Cambridge, Massachusetts 02138, USA}
\author{C.\ M.\ Marcus}
\affiliation{Department of Physics, Harvard University, Cambridge, Massachusetts 02138, USA}
\author{M.\ P.\ Hanson}
\affiliation{Materials Department, University of California, Santa Barbara, California 93106, USA}
\author{A.\ C.\ Gossard}
\affiliation{Materials Department, University of California, Santa Barbara, California 93106, USA}
\date{\today}

\begin{abstract}

We report coherent operation of a singlet-triplet qubit controlled by the arrangement of two electrons in an adjacent double quantum dot. The system we investigate consists of two pairs of capacitively coupled double quantum dots fabricated by electrostatic gates on the surface of a GaAs heterostructure.  We extract the strength of the capacitive coupling between qubit and double quantum dot and show that the present geometry allows fast conditional gate operation, opening pathways to multi-qubit control and implementation of quantum algorithms with spin qubits.
 \end{abstract}

\maketitle

Advances in control of single electrons in quantum dots \cite{hanson07} have led to the prospect of using electron spin as a quantum bit (qubit) in quantum computation \cite{loss98}. One formulation of the qubit uses singlet \mbox{\(\mid\)\(S\rangle=\frac{1}{\sqrt{2}}(\mid\uparrow\downarrow\rangle-\mid\uparrow\downarrow\rangle)\)} and triplet \mbox{\(\mid\)\(T_{0}\rangle=\frac{1}{\sqrt{2}}(\mid\uparrow\downarrow\rangle+\mid\uparrow\downarrow\rangle)\)} states \cite{levy2002} of two electrons in a double quantum dot (double QD, DQD) (Fig.~1(a)). Most requirements for quantum computing \cite{divincenzo2000} with this qubit have been met \cite{petta05,barthel09,bluhm10,barthel10}, including all electrical full single-qubit control \cite{foletti09}. Rotation about the $z$-axis of the Bloch sphere (Fig.~1(a)) is governed by the exchange interaction between two spins, which can be controlled electrostatically near degeneracies of the charge arrangement of the two electrons. Rotation about the $x$-axis is mediated by gradients of the Zeeman field, produced either by nuclear gradients \cite{foletti09} or permanent magnets \cite{pioro08}. 

The electrostatic interaction between DQDs was identified theoretically to lead to a two-qubit interaction sufficient for universal quantum computation \cite{taylor05}. In this scheme, the control (C) DQD is configured to allow its spin configuration ($S$ or $T_0$) to determine its charge state via Pauli blockade \cite{ono02} of the charge transition from the singly occupied (1,1) to the doubly occupied (0,2) (or (2,0)) configuration, where (N\(_\text{L}\),N\(_\text{R}\)) are the absolute electron occupancies of the left and right QD. That is, rapid charge relaxation into the symmetric orbital ground state of (0,2) occurs only for the spin-antisymmetric singlet ($S$) state, while the spin-symmetric triplet (\(T_{0}\)) remains trapped in the (1,1) charge configuration \cite{petta05}. The resulting charge state of the control DQD in turn influences the rate of coherent state evolution in the target (T) DQD through the dependence of the exchange interaction on electrostatic tuning.  The two-qubit operation is thus mediated by the charge configuration of the control DQD (Fig.~1(b)). Here we control the coherent operation of a singlet-triplet qubit using the directly controlled charge configuration of a second proximal DQD, providing key parameters of the capacitive two-qubit interaction.

\begin{figure}[b]
    \includegraphics[width =3.2 in]{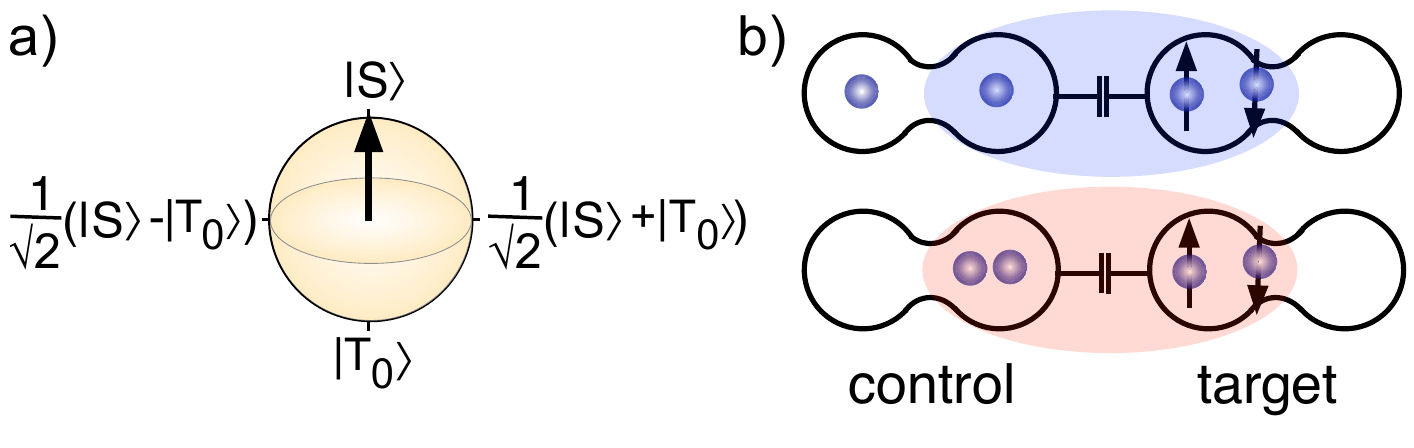}
    \caption{(Color online). (a) Bloch sphere representation of the singlet-triplet qubit, which is formed by the singlet \textit{S} and  \(m_{s} = 0\) triplet \(T_{0}\) electron spin states of two singly occupied quantum dots. (b) Electrostatic interaction between proximal double quantum dots (DQDs) alters the rate of coherent evolution in one DQD depending on the charge configuration of the other DQD.}
\end{figure}
 
\begin{figure}
    \includegraphics[width = 3.2 in]{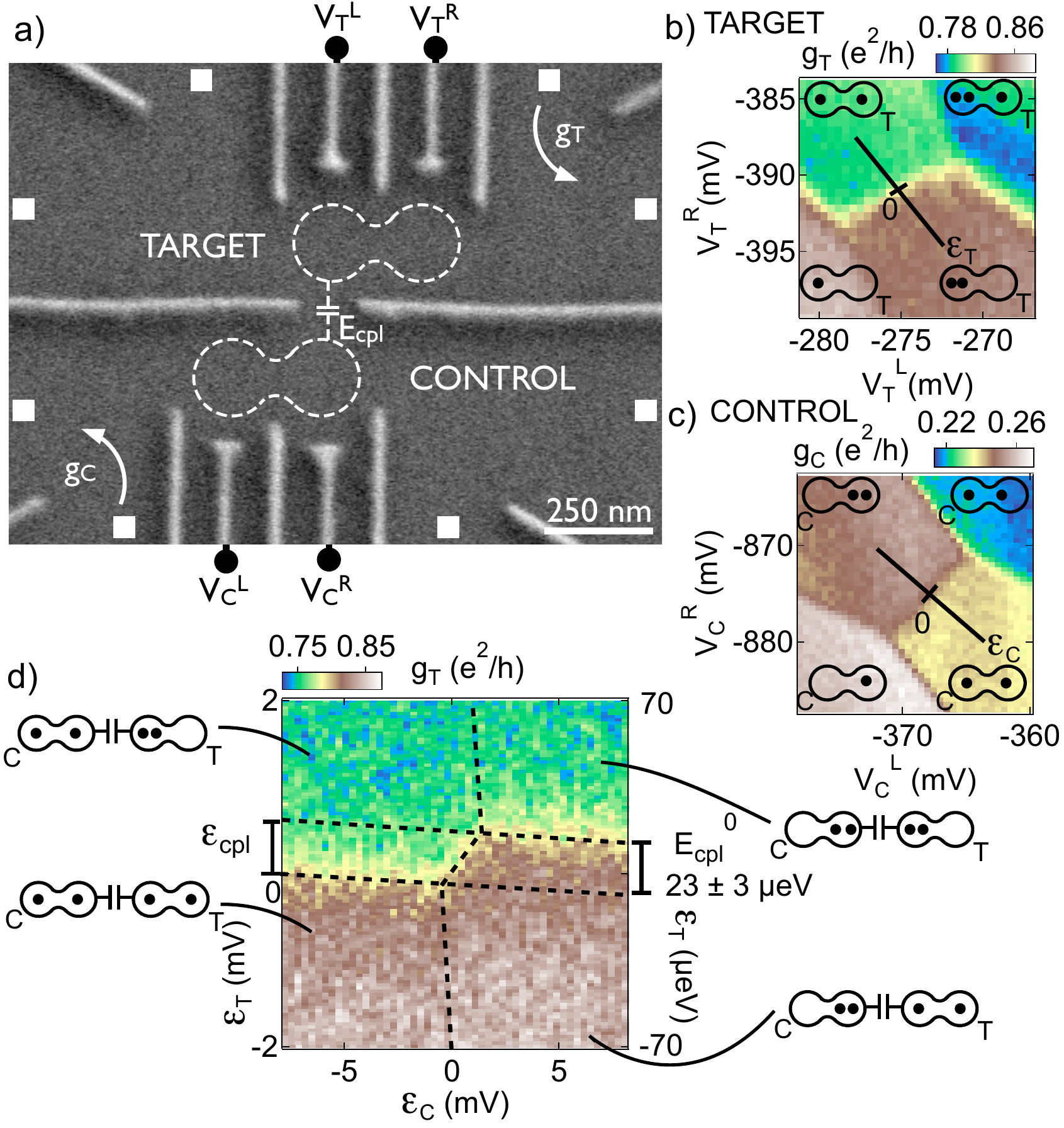}
    \caption{(Color online). (a) Micrograph of a device similar to the one measured. Gate voltages \(V_{\rm T}^{L}\) and \(V_{\rm T}^{R}\)  (\(V_{\rm C}^{L}\)  and \(V_{\rm C}^{R}\))  control the charge state of the target (control) double quantum dot (DQD). Quantum point contacts (QPCs) with conductances \(g_{\rm T}\) and \(g_{\rm C}\) sense charge states of target and control DQDs . The DQDs are capacitively coupled by an electrostatic interaction \(E_{\rm cpl}\). (b) ((c)) QPC conductance measured as a function of gate voltages \(V_{\rm T}^L\) and \(V_{\rm T}^R\) (\(V_{\rm C}^L\) and \(V_{\rm C}^R\)) dot shows distinct conductance levels \(g_{\rm T}\) (\(g_{\rm C}\)) for each electron configuration (N\(_\text{L}\),N\(_\text{R}\))\(_\text{\rm T}\) ((N\(_\text{L}\),N\(_\text{R}\))\(_\text{\rm C}\)). Detuning axes \(\epsilon_{\rm T}\) and \(\epsilon_{\rm C}\) for target and control DQD are indicated. (d) Voltage detuning, \(\epsilon_{\rm T}\), of the target DQD as a function of the voltage detuning, \(\epsilon_{\rm C}\), of the control DQD. The shift of the target detuning axis \(\epsilon_{\rm cpl}\) that occurs when the occupancy of the control DQD changes is indicated on the left axis. The right axis shows the corresponding energy shift \(E_{\rm cpl}^{0}\).}
\end{figure}

A pair of DQDs were defined with Ti/Au depletion gates on a GaAs/Al\(_\text{0.3}\)Ga\(_\text{0.7}\)As heterostructure with 2DEG 110 nm below the surface (Fig.~2(a)). 2DEG mobility was $2\times10^\text{5}$ cm\(^\text{2}\) V\(^\text{-1}\) s\(^\text{-1}\) with electron density 2\(\cdot10^\text{15}\) m\(^\text{-2}\). Electron temperature was $\sim 150$~mK. The $S=1$ triplet states were separated using an external magnetic field $B_{\rm ext} = 0.1 T$ applied in the plane of the 2DEG.   

Electron configurations in both DQDs were controlled by tuning the voltages applied to the plunger depletion gates \(V_{\rm C(T)}^{L}\) and \(V_{\rm C(T)}^{R}\), and were measured with proximal quantum point contact (QPC) sensors \cite{field93, elzerman03}. The control (target) DQD was tuned to the (1,1)\(_\text{{\rm C}}\)-(0,2)\(_\text{{\rm C}}\) ((1,1)\(_\text{{\rm T}}\)-(2,0)\(_\text{{\rm T}}\)) charge transition (Figs.~2(b) and 2(c)) where Pauli blockade was observed for both DQDs in both transport and charge sensing \cite{johnson05c}. 

Energy detuning axes \(\epsilon_{\rm C}\)  and \(\epsilon_{\rm T}\) were defined along the (1,1)\(_\text{\rm C}\)-(0,2)\(_\text{\rm C}\)  and (1,1)\(_\text{\rm T}\)-(2,0)\(_\text{\rm T}\) charge transitions of the control  and target DQDs, as shown in Figs.~2b and 2c. 

The strength of the capacitive interaction between DQDs defines a coupling strength, \(E_{\rm cpl}^{0}\), given by the energy difference between the ((0,2)\(_\text{\rm C}\)(2,0)\(_\text{\rm T}\)) charge configuration and the ((1,1)\(_\text{{\rm C}}\)(1,1)\(_\text{\rm T}\)) configuration. When the control DQD was tuned to the (0,2)\(_\text{\rm C}\) charge state, the (1,1)\(_\text{\rm T}\)-(2,0)\(_\text{\rm T}\) charge transition of the target DQD shifted to a more positive detuning by an amount \(\epsilon_{\rm cpl}\) (Fig.~2(d)). This shift in detuning reflects an increased energy of the ((0,2)\(_\text{\rm C}\)(2,0)\(_\text{\rm T}\)) state resulting from capacitive coupling between DQDs. The detuning voltage shift of 0.63 mV, when converted to energy based on finite-bias transport measurements, gives \(E_{\rm cpl}^{0}\) = 23 \(\pm\) 3~\(\mu\)eV.   

\begin{figure}[b]
    \includegraphics[width= 3.2 in]{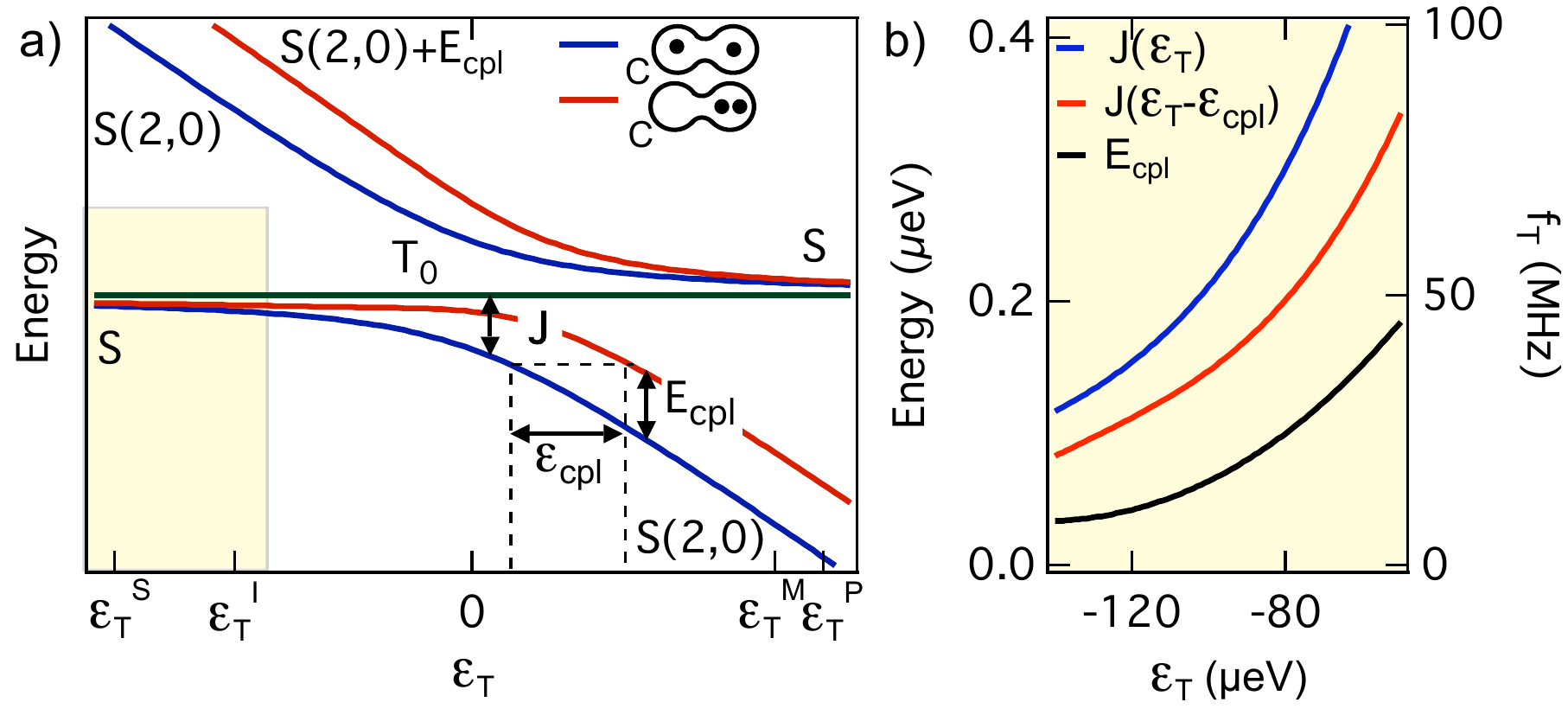}
   \caption{(Color online). (a) Energy diagram near the (1,1)\(_\text{\rm T}\) - (2,0)\(_\text{\rm T}\) transition of the target double quantum dot (DQD). Energy levels of the hybrid singlet state as a function of target detuning \(\epsilon_{\rm T}\) for (1,1)\(_\text{\rm C}\) (blue) and (0,2)\(_\text{\rm C}\) (red) occupation of the control DQD. Detuning of the target qubit at which separation of the electrons in separate quantum dots \(\epsilon_{\rm T}^{S}\), interaction of the two double quantum dots \(\epsilon_{\rm T}^I\), measurement \(\epsilon_{\rm T}^{M}\) and singlet preparation \(\epsilon_{\rm T}^{P}\) take place during coherent manipulation are indicated. The yellow area indicates detuning range considered in b. (b) Singlet-triplet energy splittings and corresponding target qubit precession frequency $f_{\rm T}$ for control double quantum dot occupation (1,1)\(_\text{\rm C}\) (blue) and (0,2)\(_\text{\rm C}\) (red). Difference in exchange energies, \(E_{\rm cpl}(\epsilon_{\rm T})\) (black) determine the duration for conditional operation.}
\end{figure}

\begin{figure}
    \includegraphics[width = 3.2 in]{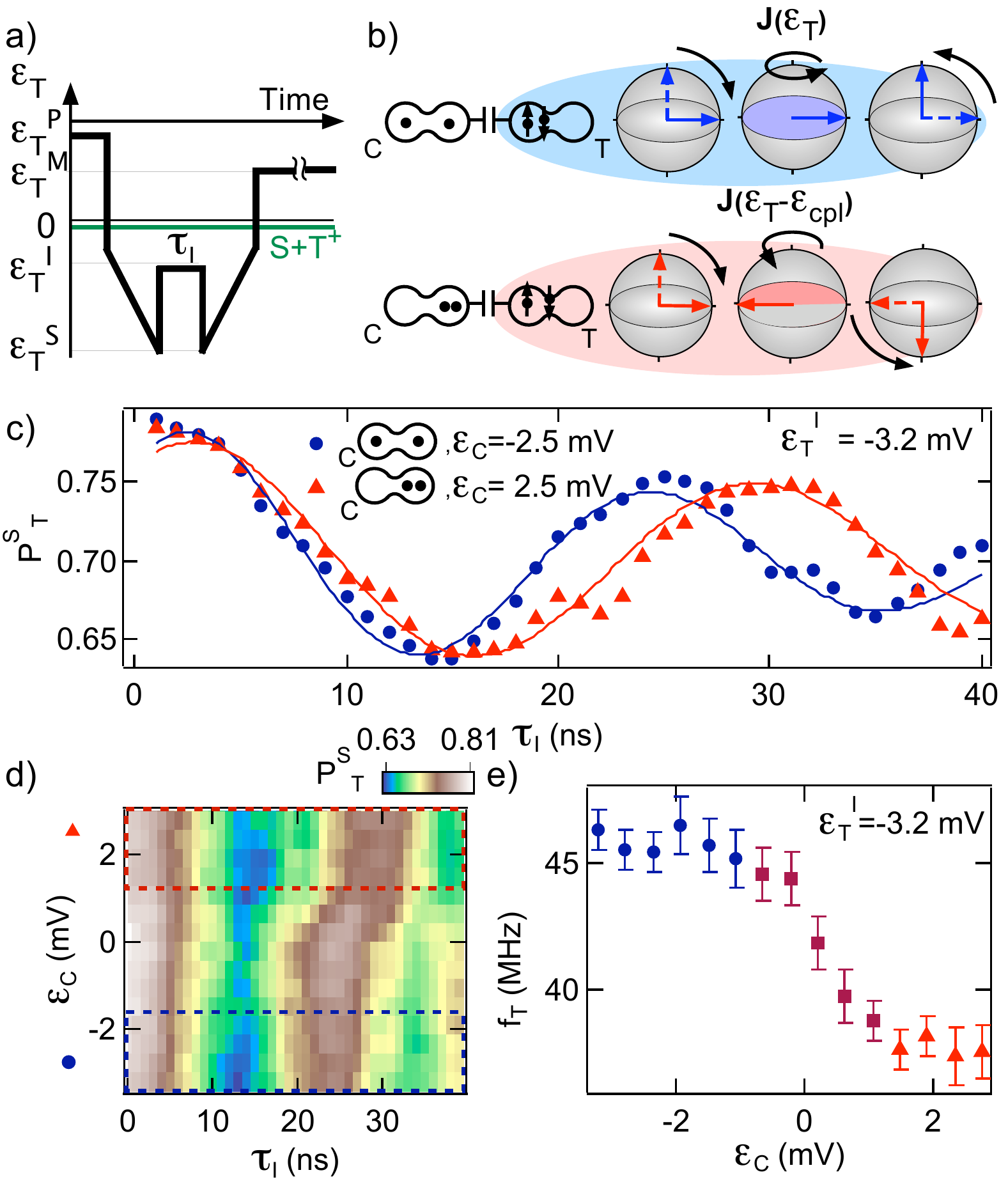}
     \caption{(Color online). (a) Pulse sequence used in coherent manipulation of the target qubit. Target detuning \(\epsilon_{\rm T}^{P}\) for singlet preparation, \(\epsilon_{\rm T}^{S}\) for adiabatic loading of the singlet-triplet superposition state, \(\epsilon_{\rm T}^{I}\) for exchange and coupling interaction and \(\epsilon_{\rm T}^{M}\) for measurement, are indicated. (b) Bloch sphere representation of the target qubit during the sequence depicts the adiabatic loading, the precession rate of the target qubit with the control double quantum dot (DQD) in either the (0,2)\(_\text{\rm C}\) or the (1,1)\(_\text{\rm C}\) electron configuration, and the adiabatic unloading after which measurement takes place. (c) Singlet probability of the target qubit $P^S_{\rm T}$ obtained from measurement of conductance \(g_{\rm T}\) as a function of interaction time \(\tau_{I}\) at control DQD detuning $\epsilon_{\rm C} = -2.5$~mV (blue dots, control DQD in (1,1)\(_\text{\rm C}\)) and \(\epsilon_{\rm C} = \)2.2 mV (red triangles, control double quantum dot in (0,2)\(_\text{\rm C}\)). Non-zero phase of the damped cosine fits at \(\tau_{I}=\) 0  is due to the rise time of the coupling voltage pulse. (d) Measurement of spin precession as a function of control DQD detuning \(\epsilon_{\rm C}\). The dashed lines indicate the detuning of the oscillations in b. A shift in oscillation period is seen around \(\epsilon_{\rm C}\) = 0 mV when the occupation of the control DQD changes. (e) Precession frequency of the target qubit $f_{\rm T}$ as function of detuning of the control double quantum dot obtained from d. All precessions took place at a coupling detuning $\epsilon_{\rm T}^I=-$3.2~mV. Standard error of the frequencies is indicated by error bars.}
\end{figure}

Coherent manipulation of the target qubit makes use of the dependence of exchange energy on detuning along the (1,1)\(_\text{\rm T}\)-(2,0)\(_\text{\rm T}\) axis. When the charge state of the control DQD changes from $(1,1)_{\rm C}$ to $(0,2)_{\rm C}$, the detuning axis of the target qubit shifts to the right by \(\epsilon_{\rm cpl}\), as shown in Fig.~3(a). When the target DQD is fully within (2,0) (i.e., large positive $\epsilon_{\rm T}$), the shift by \(\epsilon_{\rm cpl}\) results in an increase in the energy of the target state by the maximal coupling energy, \(E_{\rm cpl}^{0}\). For $\epsilon_{\rm T} < 0$, the effect of the shift on the exchange splitting is small, vanishing for large negative $\epsilon_{\rm T}$ (Fig.~3(b)). To describe the detuning-dependent coupling strength \(E_{\rm cpl}(\epsilon_\text{T}\)) we denote, following Taylor et al.~\cite{taylor05, taylor07}, the hybrid state \(\mid\)\(\tilde{S}\rangle=\cos\theta\mid\)\(S\rangle+\sin\theta\mid\)\(S(2,0)\rangle\) on the lower branch of the anticrossing, where \(\theta=\arctan{(2\kappa(\epsilon-\sqrt{4\kappa^2+\epsilon^2})^{-1})}\) is the angle parameterizing the admixture, with \(\kappa \sim 6 \mu\)eV the interdot tunnel coupling (discussed below). With the control in (0,2)\(_\text{\rm C}\) and the target at \(\theta_\text{\rm T}\), the detuning-dependent coupling strength is given by \(E_{\rm cpl}^{0}\sin^2\theta_\text{\rm T}\) \cite{taylor05}. The effect of the control DQD changing from $(1,1)_{\rm C}$ to $(0,2)_{\rm C}$ can equivalently be represented a reduction in the singlet-triplet exchange splitting from \(J(\epsilon_{\rm T})\) to  \(J(\epsilon_{\rm T}-\epsilon_{\rm cpl})\).

To demonstrate conditional evolution, the target qubit must be manipulated before and after its interaction with the control qubit using a series of voltage pulses (Fig.~4(a)) applied to the plunger gates. A Textronix AWG 520 was used for fast gate control, allowing $\sim 1$~ns pulse rise times. This time scale is fast compared to the Overhauser precession time, \(\hbar(g\mu_{B}B_{\rm nuc})^{-1}\), preventing mixing at the $S$-$T_+$ anticrossing \cite{hanson07}, but slow compared to the charge tunneling time \(\sim\hbar /\kappa\). Here \(B_{\rm nuc}\) is the Zeeman field due to nuclei in the host material. Adiabatic loading and and unloading into and out of the $x$-$y$ plane of the Bloch sphere (Fig.~1a) is done with a 0.75$\,\mu$s ramp, which is slow compared to the Overhauser precession time \cite{petta05}. A singlet \textit{S}(2,0) was prepared in the (2,0)\(_\text{\rm T}\) charge state, after which it was adiabatically loaded into the superposition \(\frac{1}{\sqrt{2}}(\mid\)\(S\rangle+\mid\)\(T_{0}\rangle\)) in (1,1)\(_\text{\rm T}\) (Fig.~4(b)). Detuning was pulsed to a negative value \(\epsilon_{\rm T}^{I}\) where the singlet and \(T_{0}\) triplet level were separated by an energy \(J(\epsilon_{\rm T}^{I})\). Precession with frequency \(h^{-1}J(\epsilon_{\rm T}^{I})\) occurred for an interaction time \(\tau_{I}\). Following adiabatic unloading, spin-dependent tunneling into (2,0) was used to determine the singlet component of the qubit $P^{S}_{\rm T}$ from an average measurement of QPC conductance over many repeated cycles.  With the control DQD in (0,2)\(_\text{\rm C}\) the precession frequency was reduced to \(h^{-1}J(\epsilon_{\rm T}^{I}-\epsilon_{\rm cpl})\), while no such reduction was observed when the control was in (1,1)\(_\text{\rm C}\). 

The oscillation of singlet probability with interaction time \(\tau_{I}\) in Fig.~4(c) demonstrates coherent precession of the target qubit. The target precessed more slowly when the occupancy of the control DQD was (0,2)\(_\text{\rm C}\) (detuning \(\epsilon_{\rm C}=2.5\) mV) than with control DQD occupancy (1,1)\(_\text{\rm C}\) \mbox{(\(\epsilon_{\rm C}=-2.5\) mV)}.  Precession frequency $f_{\rm T}$ as a function of \(\epsilon_{\rm C}\) (Figs.~4(d) and (e)) shows that the decrease occurs near \(\epsilon_{\rm C}=0\)~mV, where the charge state of the control DQD changed from (1,1)\(_\text{\rm C}\) (\(\epsilon_{\rm C}<0\)) to (0,2)\(_\text{\rm C}\) (\(\epsilon_{\rm C}>0\)). Away from \mbox{\(\epsilon_{\rm C}\)=0 mV} no noticeable change in frequency was observed, ruling out direct effects of the gate voltages \(V_{\rm C}^{L}\) and \(V_{\rm C}^{R}\) on the precession rate, which would presumably instead appear as a continuous change in precession frequency along \(\epsilon_{\rm C}\). The coupling precession, the difference in precession rate \(h^{-1}E_{\rm cpl}(\epsilon_{\rm T}^{I})\) between both control DQD configurations, constitutes a qubit operation conditional on the charge configuration of the two electrons in the control DQD.

Fig.~5(a) demonstrates a conditional phase flip in $\sim30$~ns. In that time, the target qubit rotated \(3\pi\) through \(\frac{1}{\sqrt{2}}(\mid\)\(S\rangle+\mid\)\(T_{0}\rangle\)) and \(\frac{1}{\sqrt{2}}(\mid\)\(S\rangle-\mid\)\(T_{0}\rangle\)) states with the control in (0,2)\(_\text{\rm C}\), and through \(4\pi\) with the control in (1,1)\(_\text{\rm C}\). 

We next investigated the precession frequency and gate speed of the target qubit as a function of target detuning \(\epsilon_{\rm T}^{I}\). With the control DQD in either (1,1)\(_\text{\rm C}\) or (0,2)\(_\text{\rm C}\), the precession frequency $f_{\rm T}$ was found to increase with target qubit detuning, as expected, reflecting an increase of \textit{S}(2,0) component in the hybrid singlet state with detuning (Fig.~5(b)). A fit to the measured \(E_{\rm cpl}(\epsilon_{\rm T}^{I})\) with the theoretical \(\sin^2\theta_\text{\rm T}\)-relation of coupling frequency to detuning was made. The coupling strength \(E_{\rm cpl}^0\), which was fixed in this fit, was obtained from the detuning voltage displacement of the spin precession frequencies with (0,2)\(_\text{\rm C}\) control DQD occupation with respect to the frequencies with (1,1)\(_\text{\rm C}\) control DQD occupation. Overlap was found for a shift of $-0.32$ mV of detuning voltage, corresponding to a coupling energy \(E_{\rm cpl}^{0}\) of 11 \(\mu\)eV \cite{footnote1}. Very good agreement was found with tunnel coupling \mbox{\(\kappa\)=5.6\(\pm\)0.3 \(\mu\)eV} of the target DQD as the single free parameter in the fit. 

The fastest measured time scale for conditional precession, \(\tau_{\pi}^{\rm contr}\sim\pi\hbar(E_{\rm cpl}(\epsilon_{\rm T}))^{-1}\), defined as the time for a phase difference of $\pi$ to accumulate in the target qubit for different control states, is 20-30~ns (Fig.~5b, right axis), corresponding to \(E_{\rm cpl}(\epsilon_{\rm C,T})\sim 0.01\,E_{\rm cpl}^0$. This value can be used to infer the speed of a two-qubit singlet-triplet gate, where the spin state of the control qubit with \(\theta_\text{\rm C}\) influences the spin evolution of the target qubit. In this situation the coupling strength is given by \(E_{\rm cpl}^{0}\sin^2\theta_{\text{\rm T}}\sin^2\theta_{\text{\rm C}}\), giving a timescale for the controlled phase two-qubit gate of \(\tau_{\pi}^{\rm cond} \sim \pi\hbar(E_{\rm cpl}(\epsilon_{\rm T},\epsilon_{\rm C}))^{-1}\). If both control and target qubits were operated in the range of detunings used here, this characteristic time would be \(\sim\)100 times longer than the conditional precession time we measure, giving  $\sim3\mu$s. On the other hand, operating the target and control near zero detuning, with $E_{\rm cpl}^0\sim20 \mu$eV (for the present device geometry), yields a more favorable value, $\tau_\pi^{\rm cont}$ $\sim$ 0.4~ns. Comparison with multi-echo coherence times of order 100 \(\mu\)s \cite{bluhm10, barthel10} for individual singlet-triplet qubits suggests that the coupling strength obtained with the current device geometry, operated at small detunings, should be adequate for two-qubit gate operations. A larger coupling strength is however preferable to achieve high fidelity two-qubit operations, as working at small singlet components (i.e., at more negative detuning) is expected to yield smaller dephasing errors \cite{taylor05}. Device geometries that further enhance capacitive coupling are under development currently.

\begin{figure}
    \includegraphics[width=3.2 in]{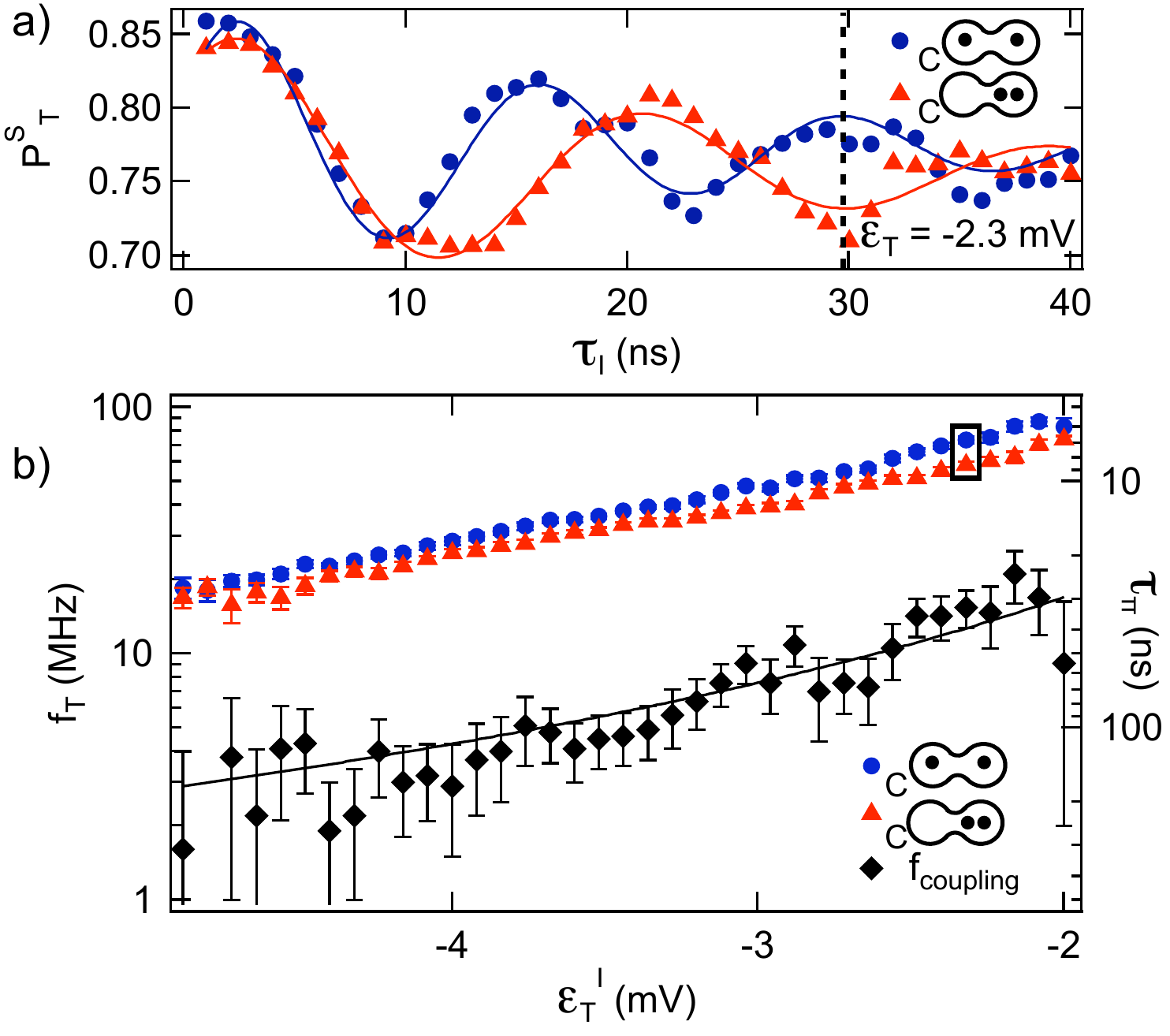}
    \caption{(Color online). (a) Singlet probability of the target qubit $P^{S}_{\rm T}$ as a function of interaction time $\tau_{I}$. After 30 ns (indicated with the dashed line) a 4\(\pi\) rotation of the target qubit has been performed when the control double quantum dot is in the (1,1)\(_\text{\rm C}\) charge state (blue dots and fit to the data), while a 3\(\pi\) rotation is performed when the control double quantum dot is in the (0,2)\(_\text{\rm C}\) charge state (red triangles and fit to the data). This corresponds to a phase flip of the target qubit conditional on the occupancy of the control double dot. (b) Precession frequency $f_{\rm T}$ as a function of target qubit detuning \(\epsilon_{\rm T}\) with the control double quantum dot in the (1,1)\(_\text{\rm C}\) (blue circles, detuning $\epsilon_{\rm C}=-8.1$ mV) and (0,2)\(_\text{\rm C}\) (red triangles, detuning \(\epsilon_{\rm C}\)=5.4 mV) charge state. Coupling frequency is the difference frequency between both data sets. Black curve is a fit to the coupling frequency data with the tunnel coupling as only free parameter. The two data points in the box correspond to the oscillations in a. The right axis shows the interaction time required for a phase flip.}
 \end{figure}

We thank T. Christian for valuable discussion and carrying out preliminary studies. This work was supported by the Intelligence Advance Research Projects Agency (IARPA) Multi-Qubit Coherent Operations (MQCO) Program, the Defence Advance Research Projects Agency (DARPA) Quantum Entanglement Science and Engineering Technologies (QuEST) Program, and the National Science Foundation (NSF) through the Materials World Network (MWN) and the Harvard  Nanoscale Science and Engineering (NSEC). Devices were fabricated at Harvard University at the Center for Nanoscale Systems (CNS), part of the NSF National Nanofabrication Infrastructure Network (NNIN).


\begin{thebibliography}{50}
\bibitem{footnote1}
	The change of coupling strength between the measurement of Fig.~1d and that of Fig.~5 is likely due to a slow drift in the quantum dot system, resulting in a change in position of the quantum dots and in their mutual coupling.
\end{thebibliography}
\end{document}